# Roman CCS White Paper

**Title:** Strongly lensed [O III] emitters at Cosmic Noon with Roman: Characterizing extreme emission line galaxies on star cluster complex scales (100 pc)

**Roman Core Community Survey:** *High Latitude Wide Area Survey*

**Scientific Categories:** *stellar populations and the interstellar medium; galaxies*

**Additional scientific keywords:** *star clusters; emission line galaxies; starburst galaxies; reionization; gravitational lensing*

**Submitting Author:**
Name: Keunho J. Kim
Affiliation: University of Cincinnati
Email: rocksteady1105@gmail.com

**List of contributing authors** (including affiliation and email):
Matthew B. Bayliss (University of Cincinnati, baylismb@ucmail.uc.edu);
Håkon Dahle (University of Oslo, hakon.dahle@astro.uio.no);
Taylor Hutchison (NASA GSFC, astro.hutchison@gmail.com);
Keren Sharon (University of Michigan, kerens@umich.edu);
Guillaume Mahler (Durham University, guillaume.mahler@durham.ac.uk);
M. Riley Owens (University of Cincinnati, m.riley.owens@gmail.com);
James E. Rhoads (NASA GSFC, jameserhoads@gmail.com);

**Overview:** Extreme emission line galaxies (EELGs) are considered the primary contributor to cosmic reionization and are valuable laboratories to study the astrophysics of massive stars. It is strongly expected that Roman's High Latitude Wide Area Survey (HLWAS) will find many strongly gravitationally lensed [OIII] emitters at Cosmic Noon (1 < z < 2.8). Roman imaging and grism spectroscopy alone will simultaneously confirm these strong lens systems and probe their interstellar medium (ISM) and stellar properties on small scales ($\lesssim$ 100 pc). Moreover, these observations will synergize with ground-based and space-based follow-up observations of the discovered lensed [O III] emitters in multi-wavelength analyses of their properties (e.g., massive stars and possible escape of ionizing radiation), spatially resolved on the scales of individual star cluster complexes. Only Roman can uniquely sample a large number of lensed [O III] emitters to study the small scale (~ 100 pc) ISM and stellar properties of these extreme emission line galaxies, detailing the key physics of massive stars and the ISM that govern cosmic reionization.

**Extreme emission line galaxies are the primary contributors to reionization and unique laboratories for the stellar feedback processes and exotic ISM conditions caused by massive stars**: EELGs are a class of galaxies with emission lines so bright that their line fluxes dominate the galaxies' light compared to the stellar continuum, as characterized by high equivalent widths (EWs) of lines such as Hα and [O III] (EW > 300 Å). The James Webb Space Telescope (JWST) has revealed that EELGs are more abundant in the early universe (z > 6) than in the present universe, which suggests that those EELGs are important contributors to cosmic reionization due to their extreme, young stellar populations (e.g., Schaerer et al. 2022; Matthee et al. 2022; Rhoads et al 2023). Thus, understanding the detailed properties of EELGs is crucial to determine how cosmic reionization occurred (e.g., how ionizing photons are generated by massive stars and escape into the Intergalactic medium (IGM)). Also, EELGs provide unique laboratories to understand the physical conditions of the ISM, including HII regions shaped by the stellar feedback of massive stars (> 20 $M_\odot$).

However, *even JWST's sharp imaging is insufficient to spatially resolve the physical properties of high-redshift (z > 6) EELGs,* which are far too small to resolve (typical effective sizes are < 0.1 arcsec). This prevents detailed investigation of the ISM and stellar properties of specific star cluster complexes; the fundamental scales that the most dense and extreme star formation operates on (e.g., Elmegreen et al. 2018; Bouwens et al. 2021; Sameie et al. 2023). Indeed, the spatially-resolved analysis of the Lyman Continuum (LyC) leaker known as the Sunburst Arc (which is strongly lensed by the foreground galaxy cluster PSZ1 G331.65-18.48), uniquely reveals that the LyC-leaking region is only a tiny part of the galaxy (radius < 50 pc, upper-limited by HST's spatial resolution; Sharon et al. 2022) with distinctly extreme physical properties that distinguish it from the non-leaking regions (Rivera-Thorsen et al. 2017, 2019; Kim et al. 2023), as shown in Figure 1. These results clearly demonstrate the importance of sufficient spatial resolution to isolate and investigate star formation conditions on star cluster scales (< ~ 100 pc) within EELGs.

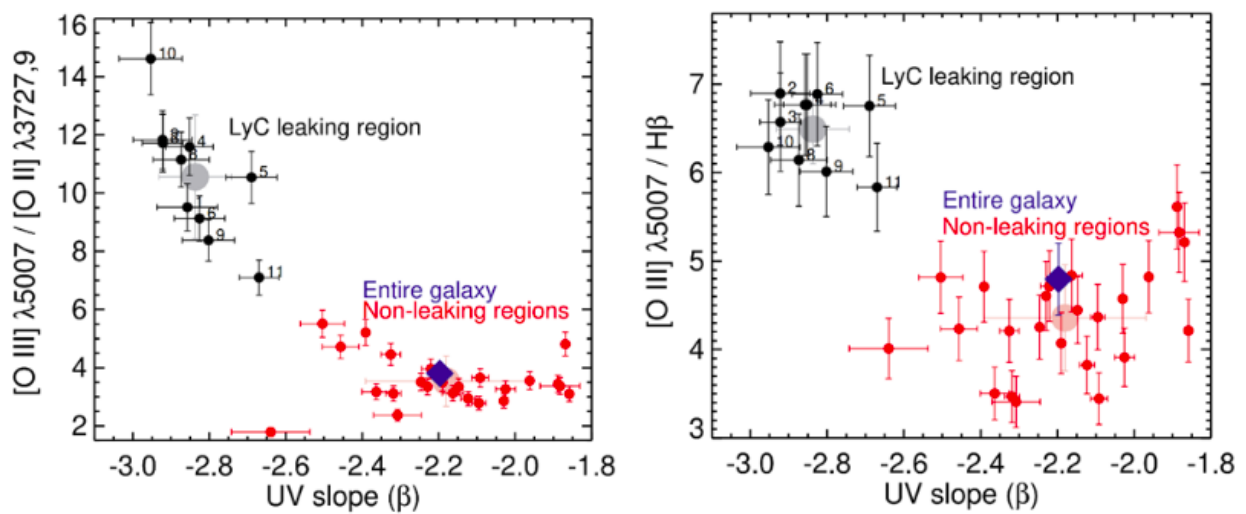

**Figure 1.** The spatially-resolved [OIII]5007/[OII]3727 (left) and [OIII]5007/Hβ (right) vs. UV-continuum slope relations reveal the distinctly extreme properties of the small (< 100 pc) LyC-

leaking region (e.g., blue UV slope and high oxygen emission line ratios) in the Sunburst Arc (PSZ1-ARC G311.65–18.48) (Rivera-Thorsen et al. 2017, 2019; Kim et al. 2023). As the only known brightly lensed LyC leaker so far, the Sunburst Arc is currently the only galaxy where such a small LyC-leaking region is clearly identified, as is necessary to robustly understand LyC escape physics. We need a large sample of such lensed LyC leakers because only lensed systems can provide detailed spatial information on scales comparable to star cluster complexes. Roman HLWAS will find many LyC candidates among lensed [O III] emitters, given that most LyC leakers are strong [OIII] emitters (Izotov et al. 2016a,b, 2018a,b; Rutkowiski et al. 2017). LyC imaging follow-up with HST will directly confirm LyC leakage, and if confirmed, spatially-resolved analysis accomplished in the Sunburst Arc would be feasible for lensed [O III] emitters.

**Strongly lensed [O III] emitters discovered by Roman:** A promising approach to achieve such detailed spatial information is to find and investigate EELGs strongly lensed by foreground massive galaxies (or galaxy clusters). With the wide survey area of 2000 $deg^2$ HLWAS is expected to identify about 400 massive galaxy clusters (> 2 *$10^{14}$ $M_\odot$), as estimated from the volume density of galaxy clusters found in the South Pole Telescope survey (Bleem et al. 2015). On a smaller scale of galaxy-scale strong lensing, 1000s of lensed sources are expected to be found out to redshift z ~ 3. While we will mainly discuss the galaxy-scale lensed systems throughout this white paper, our approach to fully leverage strongly lensed galaxies for spatially-resolved analyses is equally applicable to galaxy cluster lens systems. Galaxy-scale strong lensing occurs when a gravitational potential well of a massive foreground galaxy (typically at redshifts z = 0.3 – 0.8) magnifies the light and angular size of a distant background galaxy (typically z > 1) (e.g., SDSS SLACS, eBOSS, Bolton et al. 2006; Shu et al. 2016a,b; Cao et al. 2020, and references therein) fortuitously aligned behind the lensing galaxy. Because the strong lensing stretches out the galaxy's morphology, Roman imaging will allow a deep look at its spatially resolved morphology, often as small as star cluster complex scales of ≲ 100 pc. Figure 2 shows actual lensing systems imaged with HST with angular resolution (PSF FWHM ~ 0.09") comparable to that of Roman. The images clearly separate the foreground lens from the background source in these lens systems and reveal the distant lensed galaxies' detailed morphology. This demonstrates that Roman's imaging will both securely identify such strong lenses and provide data of sufficient quality to derive a reliable lensing mass model for each lens system.

So far, confirming galaxy-galaxy lens systems takes a lot of effort: this process typically involves first identifying the most likely lens candidates among a large data set of ground-based spectra of elliptical galaxies and finding emission lines inconsistent with the elliptical's redshift. Second, targeted follow-up HST imaging is necessary to confirm evidence of lensing by identifying any distorted, gravitationally lensed features surrounding the lensing elliptical. Due to the multi-step identification involving different ground and space-based telescopes to confirm lens systems, there are so far only tens of such lens systems confirmed with HST imaging (i.e., Bolton et al. 2006, 2008a,b; Ritondale et al. 2019). Alternatively, direct ground-based confirmation of lens systems can be possible, but typically only for systems with exceptionally large Einstein radii or during exceptional seeing conditions. Together, these conditions mean the science achievable from the ground is slow to acquire, limits the number of lens systems that can be definitively confirmed, and biases the observed population toward systems with the largest Einstein radii.

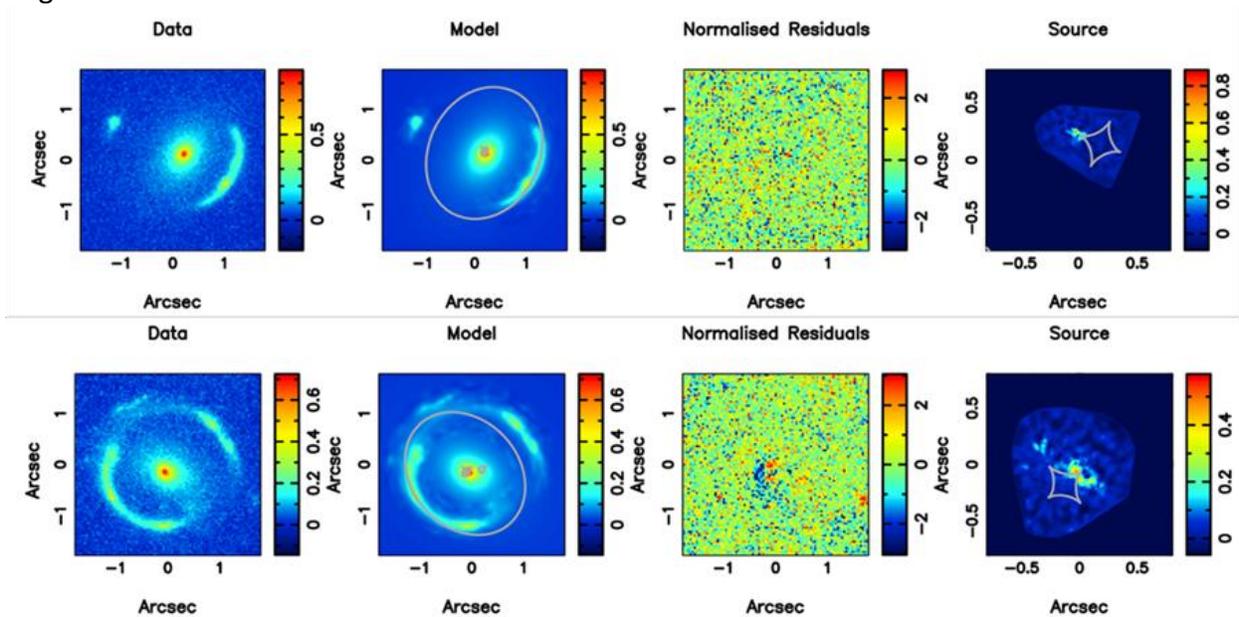

**Figure 2.** HST F606W imaging of two (top and bottom rows) galaxy-scale lens systems (lensed galaxies at 2 < z < 3) adopted from Ritondale et al. 2019. Comparable to the angular resolution of HST imaging, Roman HLWAS will image and directly confirm the lensing evidence of lensed [O III] emitters at similar redshifts. The survey's grism spectroscopy will constrain the redshifts of lensed images and foreground galaxies necessary for accurate lens modeling and magnification maps of the lens systems, as shown in the second to fourth columns of each row in this figure.

Roman will enable us to search for such lens systems very efficiently. As both imaging and grism spectroscopy are performed synchronously, we can directly confirm lensing evidence with Roman data alone. Roman imaging will reveal distinct color images of the lens system that will clearly distinguish foreground, red ellipticals and background, blue EELGs. Additionally, Roman grism spectroscopy will identify key spectral features (e.g., the [O III] and Hβ emission lines) to spectroscopically confirm background, lensed EELGs.

The High Latitude Wide Area Survey (HLWAS), combined with grism observations, will find many such galaxy-scale strong lens systems over a wide sky coverage at a superb angular resolution comparable to that of HST (i.e., PSF FWHM of 0.09 – 0.15 arcsec). Since EELGs show a color excess due to their strong emission lines (such as the [O III] 5007 line in particular), their color is dominated by the specific photometric filter within which [O III] is observed. As shown in Figure 3, this color excess identification has been successful to identify local strong [O III] emitters—nicknamed 'Green Pea' galaxies (Cardamone et al. 2009). Roman's sharp imaging and grism spectroscopy will simultaneously allow us to collect the necessary imaging and spectroscopy to robustly confirm strong lens systems.

With the currently planned four photometric filters of F106, F129, F158, and F184 for HLWAS and the grism wavelength coverage (1-2 µm), we can observe rest-frame [O III] emission over a range of redshift (1 < z < 2.8, which corresponds to the observed wavelength of [O III] within the grism wavelength coverage). Moreover, Roman data alone can constrain the spatially resolved ISM and stellar properties by mapping the ionization-sensitive [O III]/Hβ line ratio and the equivalent width distributions of [O III] and Hβ across lensed [O III] emitters, as well as deriving stellar surface density maps using the sharpest imaging, provided by the shortest wavelength filter, F106. Comparing the nebular and stellar distributions on small physical scales (~ 100 pc) within galaxies will reveal the detailed star formation physics in EELGs (e.g., whether the star formation of EELGs is mostly dominated by dense star cluster regions or by more diffuse regions).

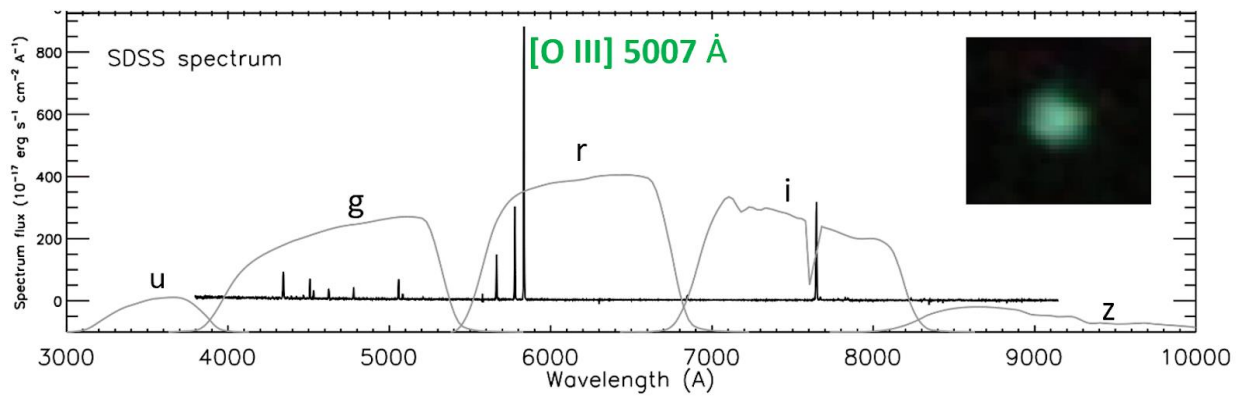

**Figure 3.** The spectrum and color-image (inset) of a strong [O III] emitter at the low redshift z = 0.16 (Cardamone et al. 2009). Due to the strong flux excess of [O III] emission (the most outstanding line in the figure) in the photometric filter, strong [O III] emitters show distinct colors, which makes them easily identifiable with their photometric colors. Similar to the distinct color of local [O III] emitters, Roman's four NIR filters for HLWAS will detect [OIII] emitters at higher redshift (1 < z < 2.8) with their distinct color, as well as lensing evidence with the foreground lensing elliptical.

**Great synergy for multi-wavelength analysis of lensed [OIII] emitters:** The discovered lensed [O III] emitters will have great synergy with multi-wavelength studies due to their redshift range (1 < z < 2.8). For instance, for a sub-sample of high-redshift galaxies (z > 2), ground-based (e.g., Gemini, Magellan, and MMT) spectroscopic follow-up can provide rest-frame UV spectroscopy, including Lyα emission. Given the close physical connections between Lyα emission and strong [O III] emission found in many (non-lensed) EELGs at all redshifts (e.g., Green Pea galaxies for low redshift, Henry et al. 2015; Yang et al. 2016, 201a,b, Izotov et al. 2020, and high-redshift z > 2, Naidu et al. 2022; Sun et al. 2023), the lensed [O III] emitters will reveal the physical connections between Lyα and [O III] emission properties in a spatially resolved manner (if accompanied by layered slit spectroscopic observations for high Einstein radius lens systems).

We can also study the production and escape of ionizing photons in promising Lyman-continuum leaking candidates among the lensed [O III] emitters, as LyC leakers are also strong

[O III] emitters (e.g., Izotov et al. 2016a,b, 2018a,b, Flury et al. 2022; Kim et al. 2023, note, however, that not all strong [O III] emitters are LyC leakers, i.e., Rivera-Thorsen et al. 2022). Lyman-continuum imaging with HST will then directly identify LyC leakers. The confirmed LyC leakers will crucially provide detailed information on the ionizing photon production and escape processes (e.g., where in a galaxy LyC is produced and escapes and whether strong [O III] emission is spatially related to the LyC leaking region(s) on sub-galactic scales), all of which are crucial information to improve our understanding of the reionization process in the early Universe. As shown in Figure 1 for the spatially resolved analysis of an only known lensed LyC leaker and EELG, the identification and characterization of such compact, ionizing star clusters is only possible with the spatial information offered by strongly lensed EELGs. With the Roman grism data, we can investigate the ionization-sensitive [O III]/Hβ line ratio distribution and identify potential LyC-leaking regions within a galaxy. For the most extreme and interesting targets, JWST imaging and spectroscopic IFU follow-up observations will provide the rest-frame optical emission line diagnostics and dust content by measuring key emission lines such as Hα, [S II], and [S III].

**Observational requirement on the HLWAS survey design:** The current observational strategy for HLWAS (i.e., four photometric filters and grism) could enable the proposed science case overall. However, our science case crucially requires multiple (3 and more) position angles for grism spectroscopy to securely distinguish the spectra from the multiple lensed images and arcs and foreground lensing galaxy expected in galaxy-scale strong lens systems. This requirement is crucial for sparse fields with many point sources in order to clearly separate the lens galaxy and lensed emitter. With such multiple position angles, deblending the grism spectroscopy would be feasible, as already demonstrated with the poorer spectral resolution HST WFC3 grism through a forward-modeling approach (e.g., Nierenberg et al. 2017).

The currently planned imaging depth (deeper than 26 mag in all bands) is likely sufficiently deep to observe the lensed galaxies, given that lensing magnified galaxies at similar redshifts (2 < z < 3) show V-mag of 23-24 with HST ACS/F606W (Ritondale et al. 2019). The current depth of grism spectroscopy for HLWAS, which is $1*10^{-16}\ erg/s/cm^2$ based on the notional exposure time of 322 s, will likely be able to measure the [O III] emission line of lensed [O III] emitters. We estimate the expected [O III] flux is on the order of 1-10 * $10^{-16}\ erg/s/cm^2$, based on the typical magnification of 17 known galaxy-scale lensed systems at similar redshifts (2 < z < 3, Ritondale et al. 2019; Cao et al. 2020) and the empirical relations of similar fluxes of Lyα emission and [O III] emission on local [O III] emitters (Yang et al. 2017a,b). However, deeper spectroscopy depth is preferable to securely detect the Hβ line, which is 3-9 times weaker given the typical [O III]/Hβ line ratio of 3-9 for EELGs (e.g., Strom et al. 2017) at similar redshifts. It is, however, likely that the Hβ emission line is still detected with the current spectroscopy depth of $10^{-16}\ erg/s/cm^2$.

**Ancillary science:** The observed images and grism spectroscopy of this proposal have practical uses for other science branches. For instance, by modeling the strong lens, the mass density profiles of the foreground early-type galaxies are measured. These mass density profiles will characterize the cold DM contents of massive early-type galaxies and their impacts on the quenching of star formation and accretion processes (e.g., Bolton et al. 2008; Shu et al. 2016a; Sharma et al. 2018).